\documentclass[journal=jpcbfk,manuscript=article]{achemso}
\usepackage{longtable}
\usepackage{multirow}
\usepackage{amsmath}
\usepackage[usenames,dvipsnames]{color}
\usepackage{soul}
\usepackage{threeparttable}
\usepackage{xr}
\usepackage{graphicx}
\usepackage{amsmath,amssymb}
\usepackage{caption}
\usepackage{color}
\usepackage{xcolor}
\usepackage{dcolumn}
\usepackage{bm}
\usepackage{float}
\usepackage{subcaption}
\usepackage{textcomp} \usepackage{url} \usepackage[normalem]{ulem}

 \usepackage{microtype}

\usepackage[unicode=true, bookmarks=true, bookmarksnumbered=true,
  bookmarksopen=true, bookmarksopenlevel=2, breaklinks=false,
  pdfborder={0 0 1}, backref=false, colorlinks=true, hidelinks
]{hyperref}

\makeatletter

\usepackage{siunitx}\usepackage{booktabs}

\makeatletter

\author{Meenu Upadhyay}\affiliation[University of Basel]{Department of
  Chemistry, University of Basel, Klingelbergstrasse 80, CH-4056
  Basel, Switzerland} \altaffiliation{Current address: NASA Ames
  Research Center, Moffett Field, California 94035, United States.}

\author{Baruch Margulis} \affiliation[Weizmann Institute of
  Science]{Department of Chemical and Biological Physics, Weizmann
  Institute of Science, 7610001 Rehovot, Israel.}
\altaffiliation{Current address: Atom Computing, 80301 Boulder, CO,
  USA}

\author{Octavio Roncero}\affiliation[CSIC]{Instituto de F\'isica
  Fundamental, CSIC, Serrano 123, 28006 Madrid, Spain}

\author{Karl Horn} \affiliation[Department of Physics]{Department of
  Physics, Technische Universit\"at Dortmund, Dortmund, Germany}

\author{Edvardas Narevicius} \affiliation[Weizmann Institute of
  Science]{Department of Chemical and Biological Physics, Weizmann
  Institute of Science, 7610001 Rehovot, Israel.}
\altaffiliation{Department of Physics, Technische Universit\"at
  Dortmund, Dortmund, Germany}

\author{Markus Meuwly} \affiliation[University of Basel]{Department of
  Chemistry, University of Basel, Klingelbergstrasse 80, CH-4056
  Basel, Switzerland}  \email{m.meuwly@unibas.ch}

\title{Quantitative Classical Relaxation Dynamics of Penning-Ionized
  He/Ne–H$^+_2$/HD$^+$ Complexes}

\begin{document}

\date{\today}

\begin{abstract}
Penning ionization allows {\it in situ} generation of ionic complexes
to launch vibrational relaxation and dissociation dynamics for rare
gas--H$_2^+$/HD$^+$ complexes. The ensuing translational kinetic
energy release (TKER) spectra characterizing the diatomic cation are a
particularly sensitive way to probe the long range part of the
intermolecular interactions. TKER distributions from quasi-classical
trajectory (QCT) simulations for the Ne--H$_2^+$ and Ne--HD$^+$
complexes quantitatively match measured positions and relative
intensities of most experimentally observed features and are also on
par with time-dependent quantum simulations. Specifically, the
QCT-simulations allow to separate signatures arising from
H$_2^+$/HD$^+$ relaxation into different final vibrational states,
such as for $(v=2) \rightarrow (v' = 0/1 )$. Trajectory-resolved QCT
analysis also provides mechanistic interpretations of the spatial and
time-dependent dynamics underlying the TKER distributions. Rare
gas--H$_2^+$/HD$^+$ complex lifetimes directly correlate with the
final rotational quantum number $j'$ of the diatomic. Low-$j'$
products are associated primarily with an initial axial approach of
the rare gas atom, whereas formation of higher-$j'$ states
predominantly results from an initial T-shaped approach geometry.
\end{abstract}

\section{Introduction}
The interaction between H$_2^+$ and rare gas atoms (Rg--H$_2^+$) is of
fundamental importance for the formation of RgH$^+$ ions in
interstellar space. For example, the ArH$^+$ was detected in the Crab
Nebula and serves as a reporter of the chemical
environment.\cite{barlow:2013} One of the essential ingredients to
characterize the ${\rm Rg} + \rm{H}_2^+ \rightarrow \rm{RgH}^+ +
\rm{H}$ reactions, which pass through the $\rm{[RgHH]}^+$ ionic
complex, is the underlying potential energy surface
(PES). Experimentally probing the $\rm{[RgHH]}^+$ intermediate is,
however, rather difficult. This is underlined by the limited number of
experimental studies available for these
systems.\cite{heh2.review:2022} Specifically, the first low-resolution
vibrational spectrum for the He--H$_2^+$ complex was only reported in
2021.\cite{asvany:2021} Up until that point the only characterization
was that of the near-dissociative states from microwave
spectroscopy.\cite{car96:395} The somewhat scarce data on Rg--H$_2^+$
complexes is even more surprising as the He--H$_2^+$ complex was
already reported for the first time in 1925.\cite{hogness:1925} For
the Ne--H$_2^+$ complex, no comparable experimental results are
available.\\

\noindent
In more recent experiments, the Rg--H$_2^+$ (Rg = He/Ne) complexes
were generated {\it in situ} through Penning ionization and the
H$_2^+$ translational spectra were characterized.\cite{MM.heh2:2023}
These spectra provide information across the entire energy range, up
to the dissociation limits of H$_2^+ (v,j)$ interacting with the rare
gas atom to query and validate rigorous computational approaches
including the underlying potential energy surface and the dynamics
method used to obtain the spectra. Based on quantum time-independent
close coupling wavepacket simulations using a reproducing kernel
Hilbert space (RKHS-)based representation of full configuration
interaction (FCI) energies it was found that the experimentally
measured resonances are reproduced within better than 10 cm$^{-1}$
over a range of $\sim 2500$ cm$^{-1}$.\cite{MM.heh2:2023} In addition,
the computed intensities of the peaks were consistent with the
measurements except for the most strongly bound H$_2 (v'=0,j'=8)$
final state.\cite{MM.heh2:2023} Using PES-morphing
techniques\cite{MM.morphing:1999} it was also demonstrated that slight
adjustments of the shape of the PES suffice to align band positions
and intensities for all observed resonances.\cite{MM.morph:2024} For
Ne--H$_2^+$ comparable experiments have been carried
out.\cite{MM.heh2:2023} Advances in molecular-beam and laser-based
experiments, coupled with improvements in the accuracy and predictive
power of atomistic simulations, have driven significant progress in
the study of state-specific and controlled molecular processes.\\

\noindent
The interaction between He and H$_2^+$ has been intensively studied in
the past using computational methods.\cite{heh2.review:2022} Among the
many PESs constructed in the past, a few are particularly notable. One
of them is based on multi reference CI calculations which was
represented as a parametrized fit.\cite{ram09:26} This PES was
successfully used in QCT and quantum dynamics studies of the
non-reactive and reactive dynamics involving the HeH$^+$ + H
asymptote. A yet more rigorous level of theory (FCI/aug-cc-pV5Z)
together with a RKHS representation was used for quantum bound state
calculations.\cite{MM.heh2:2019} This work also considered
MRCI/aug-cc-pV6Z calculations. Finally, a
correction\cite{scribano:2024} to the 2009 MRCI PES\cite{ram09:26} was
generated that removes a spurious barrier in the HeH$^+$ + H channel
to form H$_2^+$ + He. This PES was very recently used together with
the CCSD(T)-RKHS PES to thoroughly investigate the HeH$^+$ + H
$\rightarrow$ H$_2^+$ + He reaction.\cite{MM.heh:2025} For the
H$_2^+$--Ne system a comparatively small number of PESs exist. A
recent one was used in time-independent close-coupling quantum
wavepacket studies of the dissociation dynamics H$_2^+$--Ne complex to
form H$_2^+$ + Ne.\cite{MM.heh2:2023}\\

\noindent
The main focus of the present work is to determine translational
kinetic energy release (TKER) spectra of the separating H$_2^+$/HD$^+$
ions away from the Rg--H$_2^+$/HD$^+$ (Rg = He/Ne) complexes. This
process is particularly sensitive to the long-range part of the
PES. Hence, the experiments provide valuable, high-quality information
about regions of the PES that are otherwise difficult to probe
directly. Simulations require full-dimensional, non-reactive PESs of
sufficiently high quality for accurate studies. For the He--H$_2^+$
system, RKHS- and KerNN-represented PES based on
FCI/aug-cc-pV5Z\cite{MM.heh2:2019} and CCSD(T)/aug-cc-pV5Z reference
calculations are available.\cite{MM.heh:2025} For Ne--H$_2^+$, PESs
have been constructed by fitting polynomial functions to MRCI and
MRCI+Q datasets\cite{lv2010exact,xia11:1486} together with a more
recently developed RKHS-representation using CCSD(T) reference
data.\cite{MM.heh2:2023} These polynomial PESs have been used for the
Ne--H$_2^+$ $\rightarrow$ ${\rm NeH}^+ + {\rm H}$ reaction which
agreed qualitatively with measurements of the total cross sections.\\

\noindent
This work is structured as follows. First, the methods are
described. This is followed by results from QCT and two different
quantum wavepacket simulations for the Ne+H$_2^+$ complex comparing
with measurements. Next, the Ne+HD$^+$ system is discussed. Then,
spatial aspects of the dynamics are presented and the results are
discussed in a broader context. Then, conclusions are drawn.\\

\section{Methods}

\subsection{Potential Energy Surfaces}
Classical and quantum simulations were carried out using different
PESs. For He--H$_2^+$ this included the non-reactive FCI/aug-cc-pV5z
RKHS\cite{MM.heh2:2019} (``FCI") and reactive CCSD(T)/aug-cc-pV5z
KerNN\cite{MM.heh:2025} (``CCSD(T)") PESs whereas for
Ne--H$_2^+$/HD$^+$ the non-reactive CCSD(T)/aug-cc-pV5z
RKHS\cite{MM.heh2:2023} (referred to as ``CCSD(T)") PES were
employed. The motivation to consider different PESs for the same
system was to map differences in the shapes of the PESs on the
observables of interest, in particular the long-range part of the
interactions which are particularly relevant for post-Penning
dynamics. All representations use Jacobi-coordinates $(r,R,\theta)$,
where $r$ is the diatomic bond-length, $R$ is the separation between
the rare gas atom and the center-of-mass of the diatomic, and $\theta$
is the angle between the two unit vectors $\vec{r}$ and $\vec{R}$.\\

\noindent
The accuracy of the RKHS representation for the \textit{ab initio}
data has been assessed in previous
studies.\cite{MM.heh2:2019,MM.heh2:2023,MM.heh:2025,MM.neh2p:2025} The
RKHS model shows excellent agreement with the reference energies,
achieving a root mean square error (RMSE) of only a few cm$^{-1}$,
highlighting its reliability in capturing the underlying PES. At
FCI/aug-cc-pV5z level,\cite{MM.heh2:2019} the equilibrium geometry of
He--H$_2^+$ is found to be linear (He--H--H) with bond lengths $r =
2.0749$ a$_0$ and $R = 2.9712$ a$_0$, and a total energy of
$-25259.3053$ cm$^{-1}$ relative to complete fragmentation. For
comparison, the CCSD(T)/aug-cc-pV5z\cite{MM.heh:2025} method yields a
nearly identical structure with $r = 2.0757$ a$_0$, $R = 2.970$ a$_0$,
and energy $-25252.1228$ cm$^{-1}$. The equilibrium geometry of
Ne--H$_2^+$ at CCSD(T)/aug-cc-pV5z level\cite{MM.heh2:2023} is also
linear (Ne--H--H), with optimized bond lengths of $r = 2.0812$ a$_0$,
$R = 3.3351$ a$_0$, and a total energy of $-26878.3183$ cm$^{-1}$,
also relative to full fragmentation.\\

\noindent
Other available polynomial-fitted Ne-H$_2^+$ PESs, constructed at the
MRCI/aug-cc-pV5Z (``MRCI-5") and MRCI+Q (``MRCI-4")) levels of
theory,\cite{lv2010exact,xia11:1486} report the equilibrium geometry
of Ne--H$_2^+$ with optimized bond lengths of $r = 2.0734 / 2.0718$
a$_0$ and $R = 3.2860 / 3.3247$ a$_0$, respectively. The quality of
the Ne--H$_2^+$ PESs were recently assessed by comparing with cold
collision measurements.\cite{MM.neh2p:2025}\\

\subsection{Quasi Classical Trajectory Simulations}
The QCT simulations in the present work were based on earlier
established procedures.\cite{MM.rkhs:2017, MM.co2quantum.2022,
  MM.no2:2020} Therefore, only modified technical aspects are
summarized here. Experimentally during Penning
ionization,\cite{MM.heh2:2023} the collisions were launched from the
long-range part of the PES. From experiments, the distance $R_0$
between the rare gas atom and H$_2^+/$HD$^+$ before the collision was
estimated as 8 a$_0$ for He--H$_2^+$/HD$^+$ and 10 a$_0$ for
Ne--H$_2^+$/HD$^+$.\cite{MM.heh2:2023} For valid QCT simulations,
$R_0$ needs to be greater than the impact parameter $b$. In other
words, for known initial separation of the reactants ($R_0 = 8/10$
a$_0$), the impact parameter is limited to $b < 8/10$ a$_0$.\\

\noindent
As states with total $J \in [4,6]$ contribute most to the final
states, the value of $J$ was converted into an impact parameter for
given collision energy $E_{\rm col}$ using the semiclassical
expression
\begin{equation}
    b = \hbar \sqrt{\frac{(J(J+1))}{2 \mu E_{col}}}
\end{equation}
To be consistent with experiments, collision energies were generated
from a Gaussian distribution centered at 16 K (0.00137 eV or 11.12
cm$^{-1}$) with FWHM of $\pm 20$ cm$^{-1}$ (0.0024 eV or 28.7 K) and
only $E_{\rm col} \geq 0$ were retained for subsequent simulations. To
more realistically reflect the experimental conditions, the He/Ne
velocities were assigned equal probabilities for motion towards and
away from H$_2^+$/HD$^+$. This opens the possibility for trajectories
not to form a Rg--H$_2^+$/HD$^+$ complexes.\\

\noindent
For the QCT simulations, Hamilton's equations of motion were solved
using a fourth-order Runge-Kutta numerical method with a time step of
$\Delta t = 0.05$ fs, ensuring the conservation of total energy and
angular momentum throughout the dynamics. Because the final $v'$ and
$j'$ quantum numbers are real-valued rather than strictly quantized, a
rigorous selection criterion needs to be applied. Specifically, in the
analysis of the TKER spectra, only trajectories for which the
(real-valued) final $v'$ and $j'$ differ by no more than 0.01 from the
nearest integer values were considered. In addition, the analysis
accounted for the nuclear-spin symmetry selection rules of
($\mathrm{H_2}$): for para-($\mathrm{H_2}$) $(j=0)$, only even final
$j'-$states were included, whereas for ortho-($\mathrm{H_2}$) $(j=1)$,
only odd $j'-$rotational states were retained. The resulting para- and
ortho-contributions were subsequently combined using their statistical
nuclear-spin weights ($1:3$) to construct the final TKER spectra. Out
of the $2 \times 10^6$ trajectories that were run, only 0.1\% met the
criterion and were used for the TKER analysis.\\

\subsection{Quantum Simulations}
The quantum wavepacket simulations use Jacobi coordinates with
$\vec{\mathbf{r}}$ the vector between the hydrogen atoms,
$\vec{\mathbf{R}}$ the vector from the dihydrogen center of mass to
the rare gas atom and $\theta$ the angle between the two vectors. With
$R=|\vec{\mathbf{R}}|$ and $r=|\vec{\mathbf{r}}|$, the total
Hamiltonian is
\begin{equation}
    H_{\mathrm{tot}} = -\frac{\hbar^2}{2\mu_{\mathrm{cmplx}}}
    \nabla^{2}_{\vec{\mathbf{R}}} -\frac{\hbar^2}{2\mu_{\mathrm{diat}}}
      \nabla^{2}_{\vec{\mathbf{r}}} + V(R,r,\theta)\,\,,
    \label{eq:total_hamiltonian}
\end{equation}
where $\mu_{\mathrm{cmplx}}$ is the reduced mass of the three-body
complex, $\mu_{\mathrm{diat}}$ the reduced mass of the dihydrogen
molecule, and $V(R,r,\theta)$ the three-dimensional PES. The
eigenfunctions for the diatomic are $\chi_{v,j}(r)$ and channels
consist of tuples of quantum numbers $v, j$, and $\ell$, corresponding
to diatomic vibration, rotation and orbital angular momentum,
respectively. The total angular momentum,
$\vec{\mathbf{J}}_{\mathrm{tot}} = \vec{\mathbf{j}}+\vec{\mathbf{L}}$
obtained from coupling diatomic and orbital rotation, and parity
$p=(-1)^{j+\ell}$ are conserved under the Hamiltonian
(\ref{eq:total_hamiltonian}). Then, the channel- and parity-dependent
initial state $\Theta_{j\ell}^{Jp}(R)$ of the Rg--H$_2^+$ complex was
obtained by a simulation of the Penning ionization process. For this,
the eigenstates corresponding to shape resonances on the meta-stable
Rg$^*$-H$_2$ surface, were projected onto the ionic surface with an
optical potential $\Gamma(R)$, see \cite{MM.heh2:2023} for more
details. Depending on the initial $(J, j, \ell)$, the initial
separation between Ne and H$_2^+$/HD$^+$ is $\sim 10$
a$_0$.\cite{MM.heh2:2023}\\

\noindent
{\it Time-dependent quantum simulations:} The quantum time-dependent
dynamics was studied using the MadWave3
code,\cite{Roncero-delMazo-Sevillano:25} extracting the final state
flux on each individual rovibronic state of H$_2^+$. The wave packet
was represented in reactant (Ne + H$_2$) Jacobi coordinates (see
above) in a body-fixed frame, and using a helicity basis for the Euler
angles. The MadWave3 code was modified, defining the initial state in
a space-fixed representation as
\begin{eqnarray}
\label{eq:initial-space-fixed functions}
    \Phi^{Jp}_{v,j}({\bf r},{\bf R}) =\sum_\ell w^{Jp}_{j \ell} \quad{
      \chi_{vj}(r) \Theta_{j \ell}^{Jp}(R)\over r R} {\cal
      Y}^{JM}_{j\ell} ({\hat r},{\hat R}),
\end{eqnarray}
where the space-fixed functions, ${\cal Y}^{JM}_{j\ell} ({\hat
  r},{\hat R})$, are expressed in terms of the body-fixed functions
used in the wave packet calculations as
\begin{eqnarray}
    {\cal Y}^{JM}_{j\ell} ({\hat r},{\hat R}) &=&\\
    \sum_{\Omega\ge 0}
    \sqrt{2-\delta_{\Omega 0}} && (-1)^{j-\ell-J} \sqrt{2\ell+1} 
    \left(\begin{array}{ccc}
       \ell & j & J\\
       0  & \Omega & -\Omega
       \end{array}
    \right)
    W^{JMp}_{j\Omega}({\hat r}^{bf},{\hat R}^{bf}),\nonumber
\end{eqnarray}
Here, the body-fixed $z-$axis is parallel to the vector {\bf R} and
with the three atoms lying in the $x-z$ body-fixed frame. This
body-fixed frame is defined by the three Euler angle
$(\phi,\theta,\chi)$. The body-fixed functions are defined as
\begin{eqnarray}
    W^{JMp}_{j\Omega}({\hat r}^{bf},{\hat R}^{bf}) &=&\sqrt{
      {2J+1/8\pi^2}} [ 2(1+\delta_ {\Omega 0})]^{-1} \\ &&
    \left\lbrack D^{J*}_{M\Omega}(\phi,\theta,\chi) + p (-1)^{J}
    D^{J*}_{M\Omega}(\phi,\theta,\chi) \right\rbrack \quad
    Y_{j\Omega}(\gamma,0).\nonumber
\end{eqnarray}
In the calculations, all helicities $\Omega$ have been included for
each $J,p$ values.\\

\noindent
{\it Time-independent quantum simulations:} The time-independent
solution was obtained from solving the coupled channel equations of
the Schr\"odinger equation using the renormalised Numerov
method.\cite{johnson1978renormalized,gadea1997nonradiative} The total
wavefunction $\Psi(\vec{\mathbf{R}},\vec{\mathbf{r}})$ was written as
a product of $R-$, $r-$, and angularly dependent terms
\begin{equation}
    \Psi^{JMvj\ell}(\vec{\mathbf{R}},\vec{\mathbf{r}}) \propto
    \sum_{v'j'\ell'} G_{v'j'\ell'}^{Jvj\ell}(R) \chi_{v'j'}(r)
    \sum_{m_j=-j}^{j}\sum_{m_{\ell}=-\ell}^{\ell} C_{m_j
      m_{\ell}}^{JM}Y_{\ell,m_{\ell}}(\theta_{R},\varphi_{R})Y_{j,m_j}(\theta_r,\varphi_{r})\,\,,
  \label{eq:wavefunction_expansion}
\end{equation}
where $G(R)_{v'j'\ell'}^{Jvj\ell}$ specify the radial part of the wave
function connecting an entrance channel $(v',j',\ell')$ with all
energetically accessible exit channels $(v',j',\ell')$, see
Ref.\citenum{MM.heh2:2023} for more detail.  Similar to the TD method,
and unlike standard calculations, the overlap of the input wavepacket
$\Theta_{j\ell}^{Jp}(R)$ was propagated outwards to obtain
$\tilde{\Theta}_{j\ell}^{Jp}(R)$,\cite{MM.heh2:2023} from which the
channel-to-channel integral cross section
\begin{align}
    \sigma_{v'j'\ell'}^{Jvj\ell}=\frac{1}{2\pi}\sum_{v''j''\ell''}|A_{v'j'\ell';v''j''\ell''}\tilde{\Theta}_{v''j''\ell''}^{Jvj\ell}(R_{\max})|^2\,\,,
\end{align}
was obtained. Here,
\begin{align}
    \mathbf{A} =
    i2\left[\mathbf{F}(R_{\max})-\mathbf{G}(R_{\max})\mathbf{K}\right](\mathbf{1}+i\mathbf{K})^{-1}
\end{align}
is a matrix satisfying the physical boundary conditions, dependent on
diagonal matrices of regular ($\mathbf{F}$) and irregular
($\mathbf{G}$) spherical Bessel functions.\cite{MM.heh2:2023}. \\

\subsection{Experimental}
The experimental setup as used for generating He/Ne--H$_2^+$ complexes
was described in detail
elsewhere\cite{margulis2020direct,MM.heh2:2023} and the reader is
referred to the literature for beam characteristics and velocity map
images. To generate Ne--H$_2^+$/HD$^+$ complexes the relevant
information is summarized here. In short, two supersonic beams of
neutral rare gas atoms (He, Ne) and molecules (H$_2$ and HD) were
generated using two Even-Lavie valves.\cite{even2000cooling} The rare
gas atoms were excited to a metastable electronic state by electron
impact using a dielectric barrier discharge.\cite{luria2009dielectric}
The two beams collide at the center of a coincidence double velocity
map imaging (CDVMI) apparatus with a collision energy determined by
the angle between the valves and the relative velocity which is
controlled by the individual valve temperature and the gas
composition. Details about the supersonic beam properties, including
temperatures and velocities for Ne$^*$--HD$^+$ are given in Table
\ref{sitab:beam_prop}. Following the collision between the metastable
rare gas atom and H$_2$/HD, Penning Ionization (PI) reaction occurs
which results in {\it in situ} generation of the Rg--H$_2^+$/HD$^+$
complex and an electron e$^-$. The corresponding mass spectrum is
shown in Figure \ref{sifig:MS_NeHD}. Collision energies (see also
Table \ref{sitab:beam_prop} for Ne$^*$--HD$^+$) were selected to match
shape resonance conditions in which the neutral-neutral collision
dynamics is described by a small number of partial waves. Details
about the angular momentum contributions to the ionization probability
for Ne$^*$--HD are given in Figure \ref{sifig:W_NeHD}.\\

\noindent
Product electrons and ions are then accelerated and velocity imaged
using a set of electrodes held at constant potentials and an imaging
multi-channel plate coupled with a camera. Using Time-of-Flight
information, electron-ion coincidence pairs are identified. This
allows mass-selection of positive ion products and construction of
corresponding VMI images with complete linking between each data point
in both images. Utilizing electron-ion coincidence detection the
individual molecular ion VMI images are reconstructed corresponding to
the various initial values of $v$. The initial-state selected VMI
images for Ne$^*$--HD are shown in Figures \ref{sifig:VMI_NeHD} and
\ref{sifig:VMI_NeHD_per_v}. Angular integration of the initial
state-selected VMI images from Figure \ref{sifig:VMI_NeHD_per_v}
yields the TKER spectra for Ne--HD$^+$ reported in the results section
of the present work.\\

\section{Results}
The QCT simulations for the H$_2^+$--Ne and HD$^+$--Ne species
employed a RKHS-representation of CCSD(T) reference data. Additional
simulations were also carried out using published MRCI-4 and MRCI-5
PESs for comparison.\cite{he2010,ma2011}\\

\subsection{QCT and Quantum Wavepacket Simulations}
\begin{figure}[h!]
\centering \includegraphics[scale=0.7]{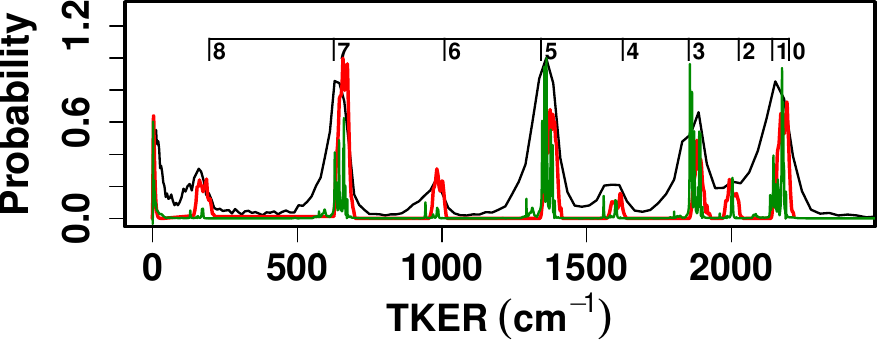}
\caption{TKER spectra for Ne + H$_2^+$ ($J=4-6, v=1, j=0/1$)
  $\longrightarrow$ Ne + H$_2^+$ ($v'=0, j'$) using QCT and QM-nuclear
  dynamics simulations using the CCSD(T)-RKHS PES. Results from QCT
  (red), TD-wavepacket simulations (green, at 2.5 K) and measurements
  (black).}
\label{fig:neh2p-v1}
\end{figure}

To set the stage, the Ne + H$_2^+$ ($J=4-6, v=1, j=0/1$) collisions
are considered first. For this process, QCT and quantum wavepacket
simulations were carried out. For the QCT simulations, the initial
separation was $R_{\rm 0}= 10$ a$_0$ and the TKER for each final state
was computed and compared with experiments in Figure
\ref{fig:neh2p-v1}. The simulated spectra were constructed by
combining the para- and ortho-H$_2^+$ contributions using the
experimental population ratio of 1:3, while enforcing the appropriate
nuclear-spin symmetry selection rules described above.\\

\noindent
Vibrational relaxation from H$_2^+$ $(v=1) \rightarrow (v'=0)$
releases sufficient energy to populate higher H$_2^+$ rotational
states $j' \in [0,8]$ as indicated in Figure \ref{fig:neh2p-v1}. This
is the origin of the measured spectra (black trace in Figure
\ref{fig:neh2p-v1}). The separation between the peaks grows according
to $\propto j'(j'+1)$. A pronounced intensity alteration for $j'$ even
(low intensity) and odd (high intensity) is evident. In the
measurements, the contributions from $j' = 0/1$ are merged into a
single peak whereas all other $j'-$components are clearly
separated. Applying the filtering procedure described in the Methods
on the QCT-trajectories, yields the red intensity pattern. It is found
that both the energies at which the measured peaks appear and their
relative intensities correctly capture the measurements. More
surprisingly, the results are also in very good agreement with
technically and computationally considerably more demanding quantum
wavepacket simulations (green). The feature at ${\rm TKER} \sim 0$
cm$^{-1}$ corresponds to trajectories that involve either elastic
collisions with H$_2^+ (v'=1)$ or dynamics in which the two collision
partner move in opposite directions which also lead to H$_2^+
(v'=1)$. In either case, this involves rather weakly bound,
long-range-dominated states of the Ne--H$_2^+$ complex. This example
suggests that a classical treatment of the process Rg + H$_2^+ (v,j)$
$\longrightarrow$ Rg + H$_2^+ (v' \leq v,j')$ for given $J$ is a
meaningful procedure.\\

\noindent
As a mutual validation, the performance of the two quantum methods was
also compared. This is done in Figures \ref{sifig:compare-wp-j0} and
\ref{sifig:compare-wp-j1}. The spectra clearly show several
overlapping resonances corresponding to highly excited X-HD$^+$
stretching excitations on different rotational states $j$. These
overlapping resonances feature small differences depending on the
quantum method (TI vs. TD) used because of differences in the two
numerical procedures. Overlapping resonances are rather sensitive to
details such as phases, and the small differences (a few cm$^{-1}$),
within the experimental resolution (ca. 10 cm$^{-1}$), are considered
to be satisfactory.\\

\noindent
Finally, the effect of using different tolerances for the
``quantization'' in the QCT-simulations was investigated. Increasing
the tolerances from 0.01 (red, the value used throughout the present
work) to 0.02 (orange), and 0.05 (fuchsia) increases the number of
trajectories that are retained. On the other hand, the probability
distributions broaden considerably, whereas the center-position of the
features remain largely unchanged, see Figure
\ref{sifig:neh2p-difftol}. Given these results, a tolerance of 0.01 is
deemed suitable, although increasing this to 0.02 would also be
possible and defensible.\\

\subsection{Simulations for Ne + HD$^+ (v=1)$ and Ne + HD$^+ (v=2)$}
\begin{figure}[H]
    \centering \includegraphics[scale=0.7]{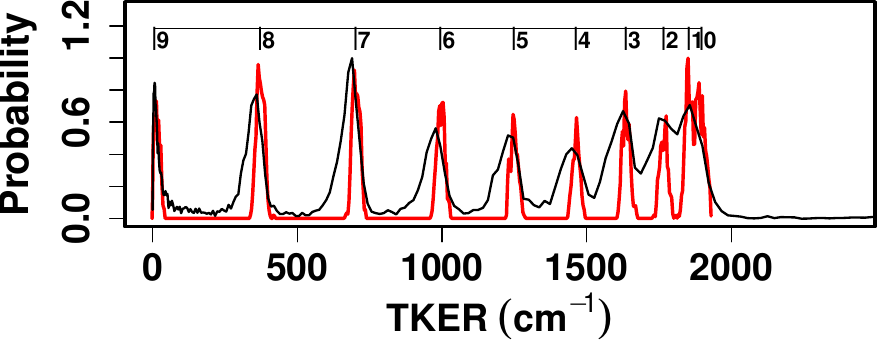}
    \caption{TKER spectra for Ne + HD$^+$ ($J=4-6, v=1, j=0$)
      $\longrightarrow$ Ne + HD$^+$ ($v'=0, j'$) from QCT simulations
      using the RKHS PES. Computed spectrum (red) compared with the
      measurements (black). The spectrum reported is for the HD$^+$
      diatom rotational level-resolved state-to-state transitions. The
      peak maximum is normalized to unity.}
    \label{fig:nehd-v1}
\end{figure}

Next, the TKER spectra for the half-collisions Ne + HD$^+ (J=4-6, v=1,
j=0)$ $\rightarrow$ Ne + HD$^+ (v'=0, j')$ and Ne + HD$^+ (J=4-6, v=2,
j=0)$ $\rightarrow$ Ne + HD$^+ (v', j')$ were considered. The QCT
simulations were carried out in the same fashion as for Ne + H$_2^+$
$(J=4-6, v=1, j=0/1)$. Starting with HD$^+ (v=1)$, only relaxation to
HD$^+ (v'=0)$ is possible, see Figure \ref{fig:nehd-v1}. Again, the
computed peak positions (red) agree rather favourably with the
measurements (black). In this case, no intensity alternations between
even and odd final $j'$ are found from the experiments. Rather, for
low and high $j'$ the intensities are high whereas for intermediate
$j' \in [4, 5, 6]$ they are comparatively reduced. Notably, the
changes in intensity distributions with H$_2^+$ and HD$^+$ as the
collision partner are also correctly described by the QCT
simulations. Similar to Ne + H$_2^+ (v=1,j)$, the 0 cm$^{-1}$ feature
is captured, including the height of the peak. In this case, the peak
at 0 cm$^{-1}$ contains contributions from HD$^+(v' = 0, j'=9)$, and
from counterpropagating fragments for which $(v=v'=1)$.\\

\noindent
The largest difference between the computations and the measurements
concerns the widths of the lineshapes. Currently the measured
linewidths are limited by the resolution of the velocity map imaging
setup used in the ion-electron coincidence spectroscopy. The
experimental resolution can be increased by a factor of 4 to 10 and
will allow a better comparison with the theory, as was recently
discussed.\cite{MM.neh2p:2025}\\

\begin{figure}[h!]
\centering \includegraphics[scale=0.7]{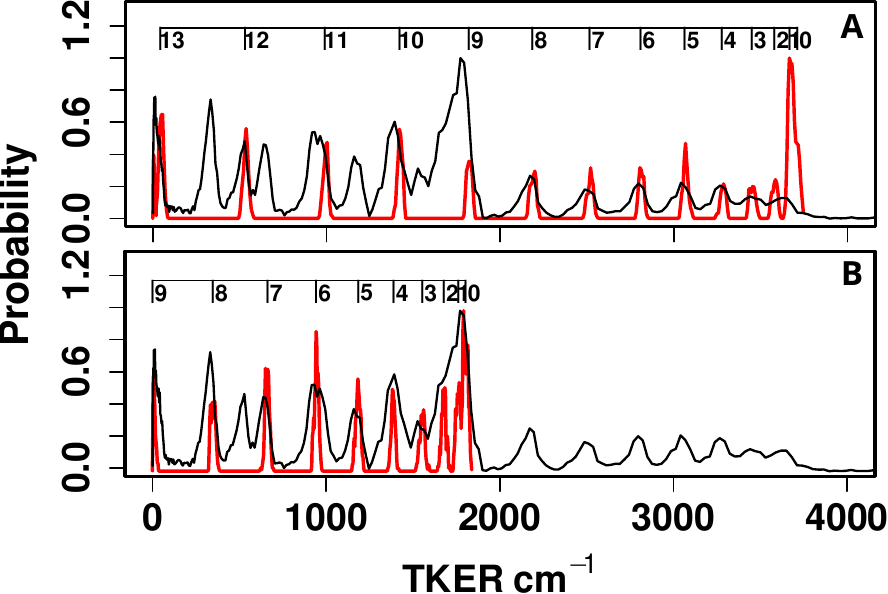}
\caption{TKER spectra for Ne + HD$^+$ ($J=4-6, v=2, j=0$)
  $\longrightarrow$ Ne + HD$^+$ ($v', j'$) from QCT simulations using
  the RKHS PES. Computed spectrum (red) compared with the measurements
  (black). The spectrum reported is for the HD$^+$ diatom rotational
  level-resolved state-to-state transitions. The peak maximum is
  normalized to unity. Panel A: for trajectories ending up in $v'=0$
  and Panel B: for trajectories ending up in $v'=1$.}
\label{fig:nehd-v2}
\end{figure}

\noindent
Results for the relaxation of Ne + HD$^+ (J=4-6, v=2, j=0)$ to Ne +
HD$^+ (v', j')$ are reported in Figure \ref{fig:nehd-v2}. In this
case, two families of lines emerge: one from relaxation to HD$^+
(v'=0,j')$, and another one to yield HD$^+ (v'=1,j')$. These are shown
separately in Figure \ref{fig:nehd-v2}A and B, respectively. The TKER
to form HD$^+ (v'=1,j')$ is qualitatively correctly reproduced. In
particular, the positions of the spectral features line up rather well
with the measurements and correctly distinguish between final states
HD$^+ (v'=1,j')$ and HD$^+ (v'=0,j')$. With regards to the
intensities, there are no particular patterns discernible from the
experiments which is also captured by the simulations except for HD$^+
(v'=1,j'=6)$ which is overpopulated compared with neighbouring peaks
and also compared with the measurements.\\

\noindent
Formation of HD$^+ (v'=0,j')$ leads to a considerably larger number of
spectral features since vibrational relaxation HD$^+ (v=2,j)$
$\rightarrow$ HD$^+ (v'=0,j')$ releases a larger amount of energy, see
Figure \ref{fig:nehd-v2}A. The measured intensity pattern for
relaxation to HD$^+ (v'=0,j' \in [0,8])$ is essentially flat, which is
largely reproduced by the simulations except for HD$^+ (v'=0,j' =
5)$. The one incorrectly described feature concerns the highest-energy
components associated with HD$^+ (v'=0,j' \in [0,1])$. The final state
HD$^+ (v'=0,j' = 9)$ is hidden under the prominent feature for HD$^+
(v'=1,j' \in [0,1])$. In the measurements, this can be seen as a
high-energy shoulder of the broad feature slightly below 2000
cm$^{-1}$.\\

\noindent
It is important to stress that experiments measure a single TKER
spectrum and assignment of the individual peaks may require additional
information. This can, for example, be obtained from QCT
simulations. For final states HD$^+ (v'=0,j' > 9)$ the measured
intensities are approximately twice as large as for HD$^+ (v'=0,j' <
9)$, which is also correctly reproduced by the QCT simulations. Even
the double-peak structure of the lowest-energy feature around 0
cm$^{-1}$ is captured. It should be noted, that this feature is not
associated with a single final state but contains overlapping
contributions. Specifically, in Figure \ref{fig:nehd-v2}A it consists
of the HD$^+(v' = 0, j' = 13)$ state and HD$^+$ ions in the $v=v'=2$
state arising from both elastic collisions and trajectories with
counterpropagating Ne and HD$^+$, whereas in panel B it consists of
the HD$^+(v' = 1, j' = 9)$ state together with the same HD$^+$
contribution. Also, the intensity of the H$_2^+$ ($v=v'=2$)
contribution at 0~cm$^{-1}$ differs between panels A and B because it
is scaled according to the population ratio of the final vibrational
channels. Specifically, in panel A it is determined by the population
ratio of the $v=v'=2$ and $v'=1$ channels, whereas in panel B it is
determined by the population ratio of the $v=v'=2$ and $v'=0$
channels.\\

\subsection{Simulations using literature PESs}
It is also of interest to carry out QCT simulations using previous
PESs for the Ne--H$_2^+$ interactions. These were generated at the
MRCI-level of theory and represented as parametrized
functions\cite{he2010,ma2011} instead of a reproducing kernel Hilbert
space. All three PESs (CCSD(T), MRCI-4, and MRCI-5) were recently used
in a comparative study based on quantum wavepacket
simulations.\cite{MM.neh2p:2025} For comparing the performance of the
three PESs, a figure of merit $\mathcal{F}$ was considered which
quantifies the difference between measured and computed peak positions
and intensities. The magnitude of $\mathcal{F}$ for the CCSD(T)-PES
was found to be lower by 5 \% to 20 \% compared with the two
MRCI-based PESs which was mainly due to an improved description of the
long-range part of the intermolecular
interactions.\cite{MM.neh2p:2025}\\

\begin{figure}[H]
    \centering
    \includegraphics[scale=0.73]{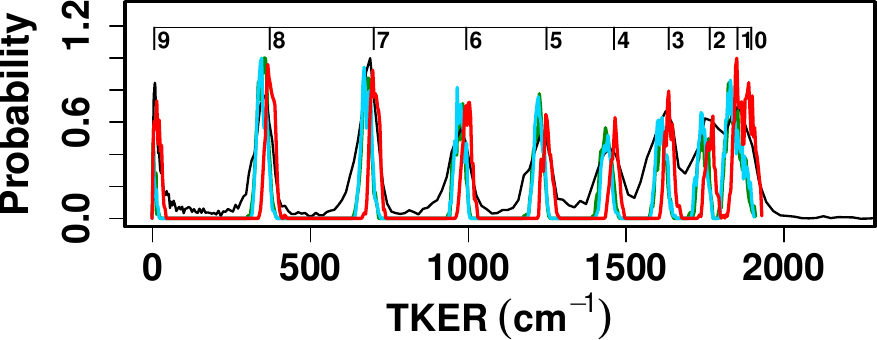}
    \caption{TKER spectra for Ne + HD$^+$ ($J=4-6, v=1, j=0$)
      $\longrightarrow$ Ne + HD$^+$ ($v'=0, j'$) from QCT simulations
      using different PESs: the CCSD(T)-RKHS PES (red),
      MRCI+Q\cite{he2010} (aqua) and MRCI-5\cite{ma2011} (green),
      compared with measurements (black). The peak maximum is
      normalized to unity.}
    \label{fig:comp-pes}
\end{figure}

\noindent
The performance of the three PESs was compared for the relaxation of
Ne + HD$^+ (J=4-6, v=1, j=0)$ $\rightarrow$ Ne + HD$^+ (v'=0, j'$),
see Figure \ref{fig:comp-pes}. TKERs from QCT simulations using the
MRCI-5 and MRCI+Q PESs are the green and aqua traces,
respectively. Both spectra overlap rather closely with one another and
also the intensity distributions are almost identical. Hence, they are
not compared separately with one another. For the final states HD$^+
(v'=0, j' \in [0,3])$, QCT simulations using the two MRCI-based PESs
lead to spectral features displaced slightly towards lower energies
compared with the measurements and the relative intensities agree well
with experiment. For HD$^+ (v'=0, j' \in [4,6])$, the peak positions
match those from experiment rather closely, whereas for HD$^+ (v'=0,
j' > 6)$ they are again displaced towards the red side of the
measurements. Contrary to that, using the RKHS PES captures final
states HD$^+ (v'=0, j' \in [0,3])$ more realistically, but features
peaks displaced to the blue for final states HD$^+ (v'=0, j' >
3)$. The largest difference, however, concerns the HD$^+ (v'=0, j' =
9)$ feature close to an energy of 0 cm$^{-1}$. QCT simulations using
the two MRCI-based PESs find a small feature, whereas using the
RKHS PES displays a line with an intensity consistent with
experiment. Vibrational relaxation and dissociation to Ne + HD$^+
(v'=0, j' = 9)$ involves trajectories which probe the long range part
and the entire angular range of the PES and do not sample the minimum
energy structure extensively, see Figure
\ref{sifig:nehdp-mapping}. Hence, the RKHS-based PES provides a
superior description of the long-range interaction between Ne and
H$_2^+$ compared with the two MRCI-based PESs.\\

\subsection{Simulations for Ne + H$_2^+ (v=2)$}
\begin{figure}[H]
    \centering \includegraphics[scale=0.7]{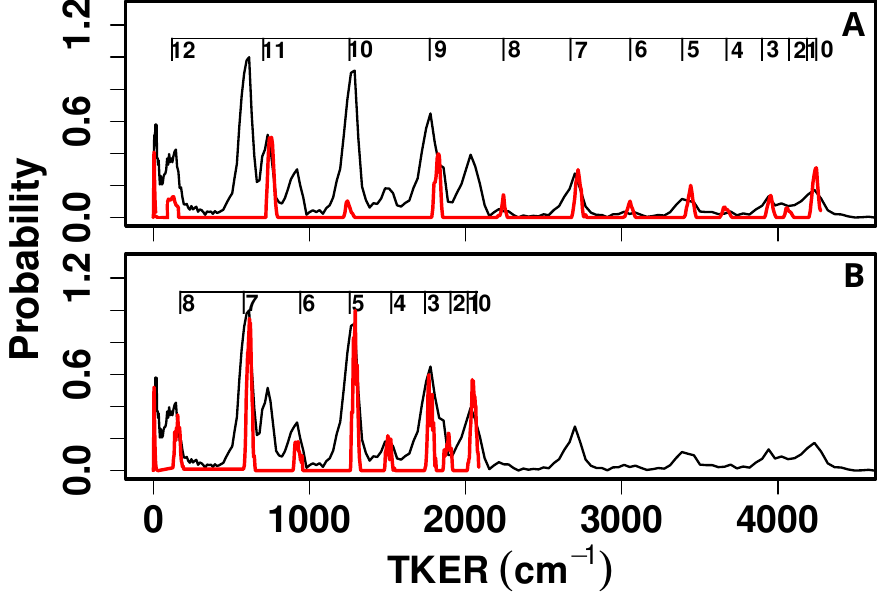}
    \caption{TKER spectra for Ne + H$_2^+$ ($J=4-6, v=2, j=0$)
      $\longrightarrow$ Ne + H$_2^+$ ($v', j'$). Comparison between
      computed (red) and measured (black) product H$_2^+$ diatom
      rotational level-resolved state-to-state averaged
      distributions. Panel A: for trajectories with final $v'=0$ and
      Panel B: for trajectories with final $v'=1$. The maximum is set
      to unity. Note: the raw peak positions are reported here and no
      scaling is used.}
       \label{fig:neh2p-v2}
\end{figure}

Finally, the TKER spectra for Ne + H$_2^+$ ($J=4-6, v=2, j=0$) to form
Ne + H$_2^+$ ($v'=0, j'$) and Ne + H$_2^+$ ($v' = 1, j'$) are
presented and discussed, see Figure \ref{fig:neh2p-v2}. For relaxation
into the vibrational ground state, see panel A, the computed spectrum
shows again a pronounced intensity alteration as was already seen for
Ne + H$_2^+$ ($J=4-6, v=1, j=0$). This is consistent with the
measurements (black trace). The region between 0 and 2000 cm$^{-1}$ is
rather congested and the QCT simulations provide a convincing
assignment of the double-peaks (around 700 cm$^{-1}$), shoulders
(around 1800 cm$^{-1}$), or overlapping features (200 cm$^{-1}$ and
1200 cm$^{-1}$) to disentangle formation of final H$_2^+$ ($v'=0, j'$)
or H$_2^+$ ($v' = 1, j'$). The peak at 0 cm$^{-1}$ here has
contributions from H$_2^+$ in $v=v'=2$ state. \\

\subsection{Spatial Analysis of the Trajectories}
\begin{figure}[h!]
    \centering
    \includegraphics[scale=0.45]{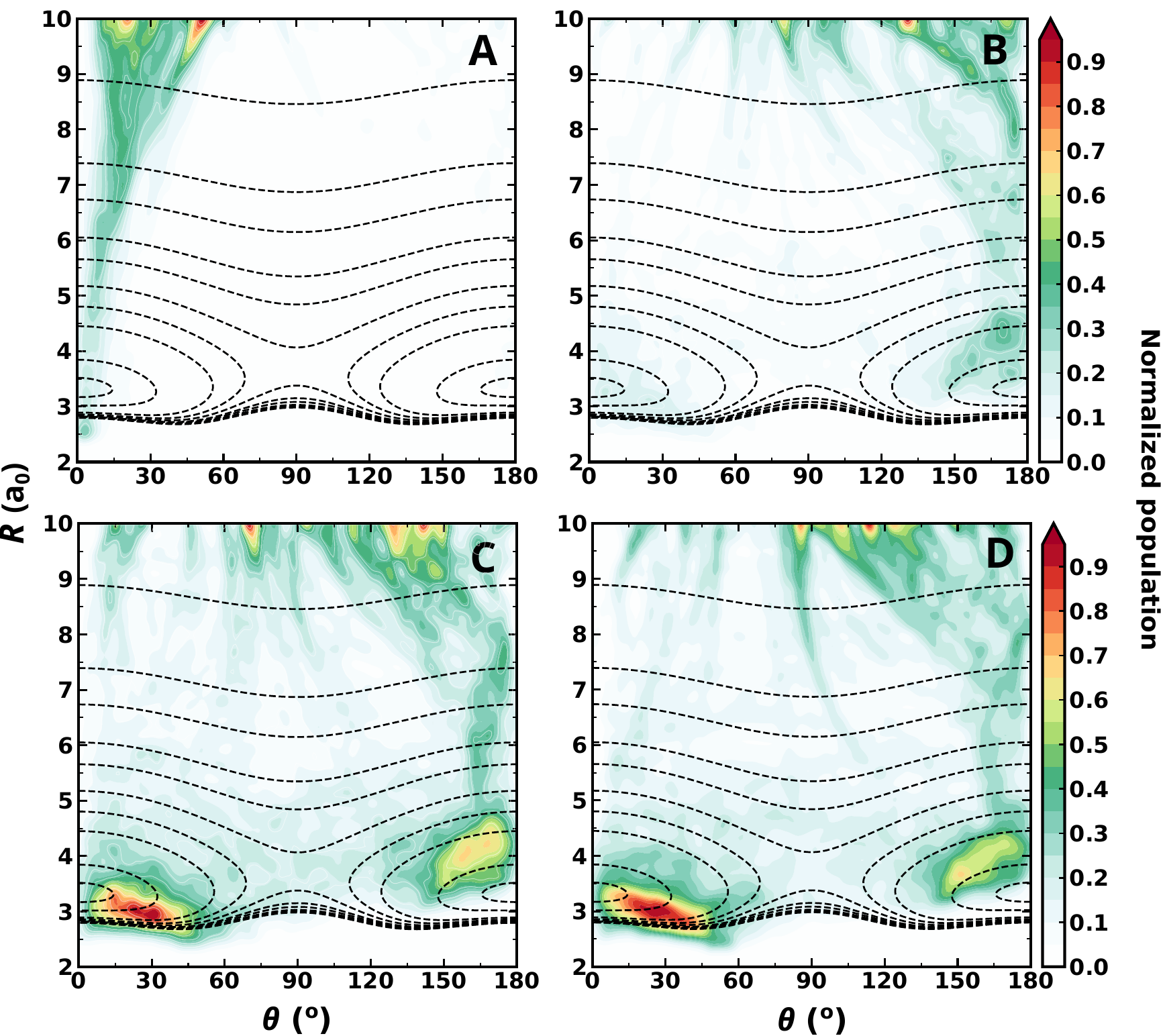}
    \caption{Density maps $P(R,\theta)$ from 100 trajectories for
      Ne+HD$^+ (v=1, j=0; J=6)$ $\rightarrow$ Ne+HD$^+ (v'=0, j'; J')$
      for $R_0 = 10$ a$_0$. The color scale represents the normalized
      probability density $P(R,\theta)$, where colors from yellow to
      red indicate regions sampled by a larger fraction of
      trajectories. The dashed isocontours represent the PES. Panels A
      to D show $P(R,\theta)$ for vibrational relaxation to different
      final states from initial HD$^+ (v=1, j=0)$: Panel A
      $\rightarrow (v'=0, j'=0)$; Panel B $\rightarrow(v'=0, j'=2)$;
      Panel C $\rightarrow(v'=0, j'=5)$; Panel D $\rightarrow(v'=0,
      j'=7)$. Here, $\theta = 0^\circ$ corresponds to the HD$^+$--Ne
      configuration, whereas $\theta = 180^\circ$ corresponds to the
      DH$^+$--Ne orientation.}
    \label{fig:map_v1tov0}
\end{figure}

One particularly attractive feature of QCT simulations is that they
provide explicit time-dependent trajectories. These trajectories can
be visualized directly or time-dependent spatial distributions can be
constructed for subsets of trajectories sharing specific
characteristics. Such analyses can provide an interpretive basis for
more abstract observables, such as peak positions in a TKER
spectrum. Figure \ref{fig:map_v1tov0} shows how vibrationally relaxing
trajectories leading to different final $v'$ and $j'$ states sample
distinct regions of the interaction potential for Ne + HD$^+$
collisions. This suggests that the trajectories corresponding to
higher $j'$ states spend more time in the deep potential well region
than those leading to lower $j'$ states. Also, the density plots
indicate that formation of a long-lived, tightly bound collision
complex is likely to produce HD$^+$ products with higher $j'$.\\

\begin{figure}[h!]
    \centering
    \includegraphics[width=0.8\linewidth]{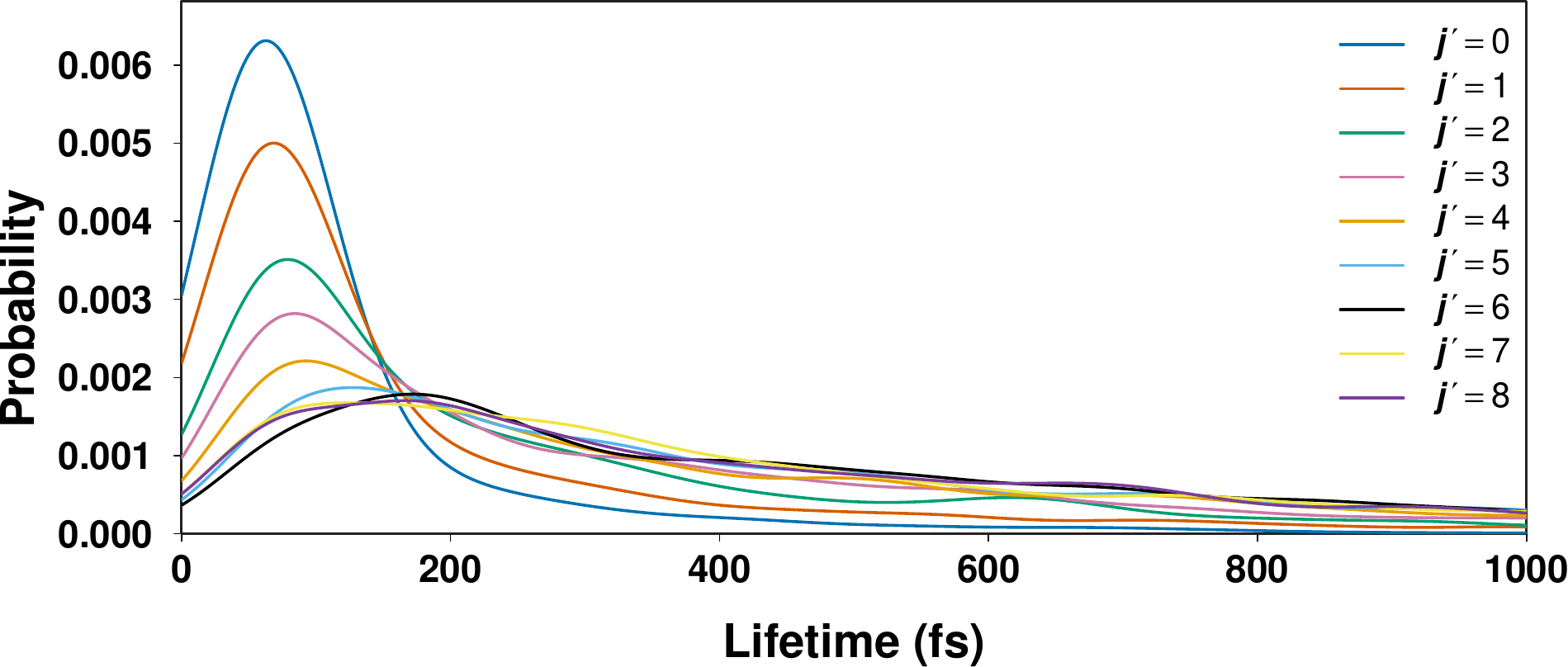}
    \caption{Normalized lifetime distributions of the collision
      complex for the Ne + HD$^+$ ($J=6$, $v=1$, $j=0$) $\rightarrow$
      Ne + HD$^+$ ($v'=0$, $j'$) process from QCT simulations on the
      CCSD(T)-RKHS PES. The lifetime is defined as the time elapsed
      between the first approach to within $R = 6$ a$_0$ and the final
      separation beyond 6 a$_0$. Each distribution is based on 1000
      trajectories for the corresponding final rotational state $j'$.}
    \label{fig:lifetime-nehdp}
\end{figure}

\noindent
The lifetime distributions for the [Ne--HD$^+$] complex for
trajectories showing vibrational relaxation, i.e., Ne + HD$^+$ ($J=6$,
$v=1$, $j=0$) $\longrightarrow$ Ne + HD$^+$ ($v'=0$, $j'$) are
reported Figure \ref{fig:lifetime-nehdp}. Here, the lifetime is
defined as the time elapsed from the first instance at which $R < 6$
a$_0$ is reached until the moment when the rare gas atom is
dissociating from the ion for which $R > 6$ a$_0$ was used. Depending
on the final $j'$, a clear shift in the peak maxima is observed:
rotationally excited products are characterized by longer median
lifetimes. \\

\begin{figure}[h!]
    \centering
    \includegraphics[width=0.6\linewidth]{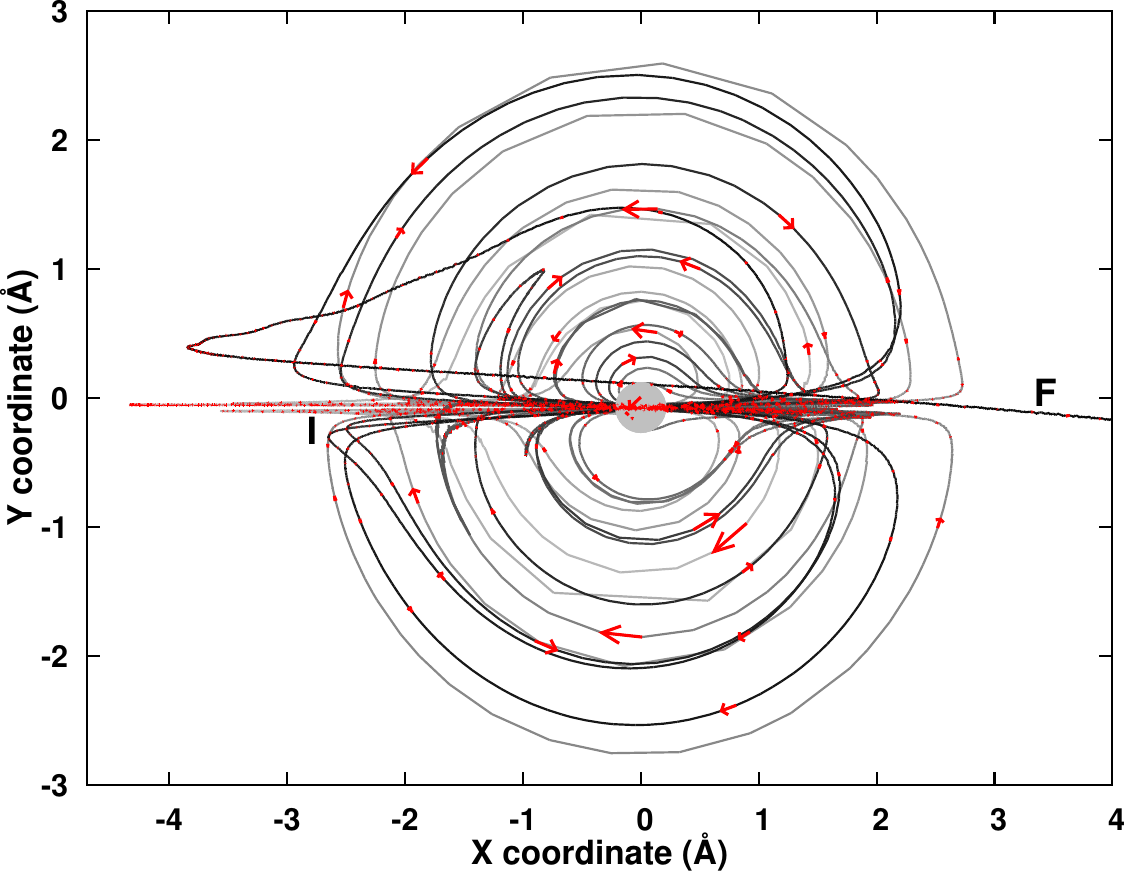}
    \caption{The motion of Ne (line with arrows) around H$_2^+$ (grey
      circle) in the $xy-$plane for the longest-living trajectory. The
      color in the plot transitions from light to dark as time
      progresses. The red arrows indicate the direction of motion and
      the length of the arrow is proportional to the average speed of
      the atom over that interval. ``I" and ``F" represent the start
      and end of the trajectory. The trajectory shown corresponds to
      the process $(v=1,j=0) \rightarrow (v'=0,j'=1)$ and the complex
      lifetime is $\sim 2$ ps.}
    \label{fig:rotation}
\end{figure}

\noindent
To characterize the rotational motion of the Ne atom around the
H$_2^+$ ion, a representative trajectory was analyzed by fixing the
center of mass of H$_2^+$ at the origin and tracking the motion of the
Ne atom relative to it. The resulting trajectory is shown in Figure
\ref{fig:rotation}. The Ne atom follows a series of semi-circular
paths, alternating between clockwise and counterclockwise motion. The
red arrows in the figure indicate the direction of motion, while their
lengths are proportional to the average velocity of the atom over each
interval, highlighting regions of accelerated and slower motion.\\

\begin{figure}[H]
    \centering \includegraphics[scale=0.57]{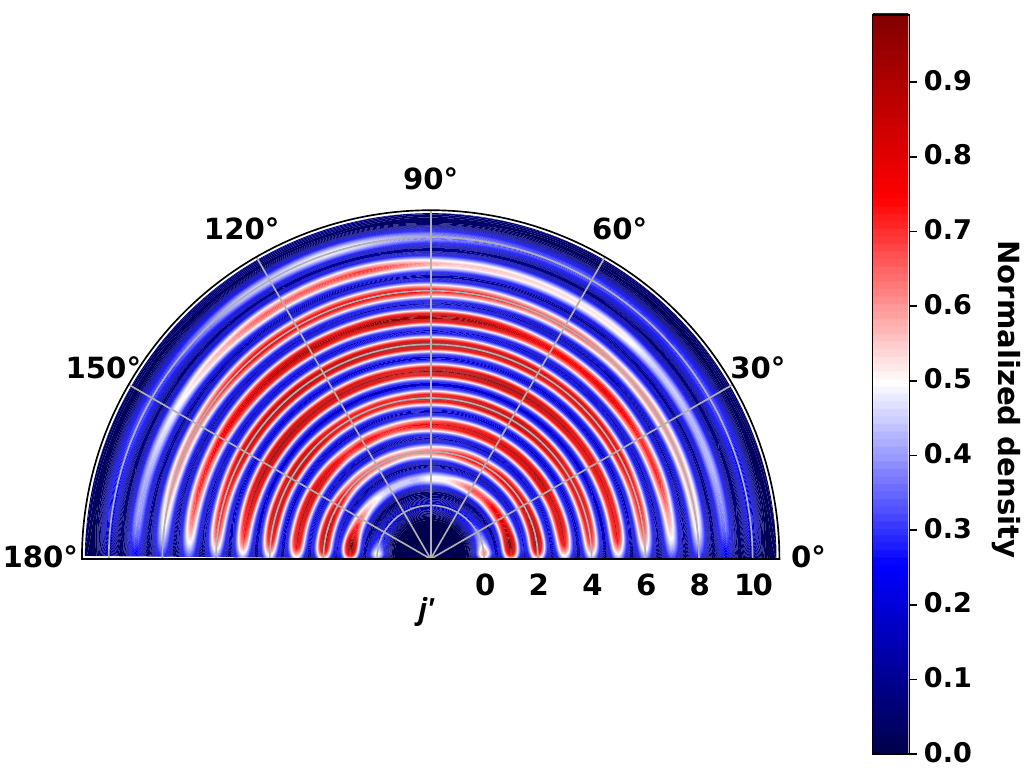}
    \caption{The probability distribution $P(\theta,j')$ to observe
      final $j'$ as a function of initial incident collision angle
      $\theta$ for Ne + HD$^+ (v=1, j=0)$ $\longrightarrow$ Ne + HD$^+
      (v'=0, j')$. The angle $\theta$ was determined at initial
      separation $R_0 = 10$ a$_0$. All colors indicate density
      relative to the maximum peak height. It is observed that final
      states with low $j' \in [0,1]$ are associated with particular
      ranges for initial incident angles $\theta \in [0, 60]^\circ$
      and $150-180^\circ$, whereas final states with $j' \in [9,10]$
      are preferred at $\theta \in [90,140]^\circ$. The asymmetry in
      the distributions with respect to $\theta = 90^\circ$ is a
      consequence of the asymmetry of the diatomic.}
       \label{fig:theta_dep}
\end{figure}

\noindent
As an alternative to sampling the impact parameter $b$, trajectories
were initialized with a specific angle $\theta$ between the incoming
rare gas atom and the axis of the diatomic HD$^+$, in order to
understand the effect of initial angular dependence on the final $j$
state of the HD$^+$. To explore this, a total of 10$^{6}$ trajectories
were initiated for initial $\theta \in [0, 180]^\circ$, with $\Delta
\theta = 0.01$, $v = 1, j=0, E_{col}=16$ K (0.00137 eV), $R_0 = 10$
a$_0$. A flat distribution of $P(\theta)$ was used to uniformly sample
the initial orientation of the Ne and HD$^+$ diatom. The results in
Figure \ref{fig:theta_dep} show a clear dependence of the final
rotational state distribution on the initial incident
angle. Specifically, low-lying rotational states $j' \in [0,1]$ are
favored at for small or large incident angles $\theta \in [0,
  60]^\circ$, In contrast, higher rotational excitations, $j' \in
[9,10]$ are preferred at intermediate angles, particularly $\theta \in
[90,140]^\circ$. At the mechanistic level, nearly collinear approaches
exert a smaller torque and therefore preferentially populate lower
rotational states whereas intermediate-angle collisions typically
exert a larger torque on HD$^+$, leading to more efficient conversion
of translational energy into rotational excitation.\\

\section{Discussion and Conclusions}
The present study finds that QCT simulations performed on accurate
PESs provide a computationally efficient and accurate approach for
describing the TKER for breakup of the Rg--H$_2^+$/HD$^+$ complexes
formed after Penning ionization. The QCT simulations quantitatively
reproduce the dominant state-to-state product distributions when
compared to both, measurements and rigorous quantum wavepacket
simulations. In addition, QCT simulations have the advantage that the
underlying (state-to-state) dynamics can be readily characterized and
investigated.\\

\begin{figure}[H]
    \centering \includegraphics[scale=0.99]{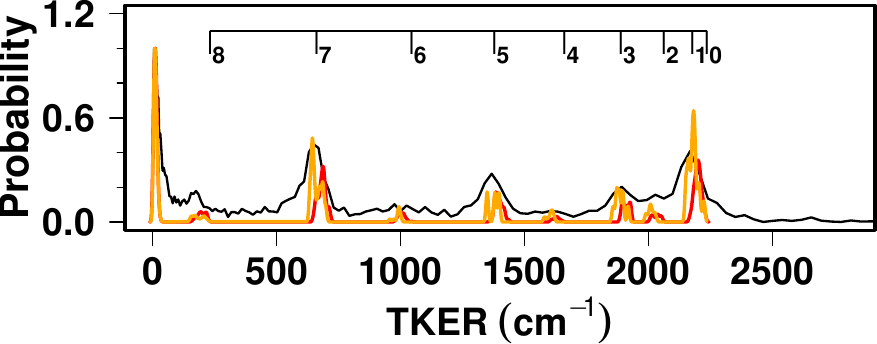}
    \caption{TKER distribution for the reaction He + H$_2^+$
      ($J=4$–$6$, $v=1$, $j=0/1$) $\longrightarrow$ He + H$_2^+$
      ($v'=0$, $j'$). Results are from QCT simulations using the
      FCI-RKHS (red) and CCSD(T)-KerNN (orange) PESs, and compared
      with the measurements (black).}
    \label{fig:heh2-v1}
\end{figure}

\noindent
This utility is further exemplified for H$_2^+$--He, see Figure
\ref{fig:heh2-v1}. For this system, QCT simulations were carried out
using a nonreactive FCI-RKHS PES (red) and a reactive CCSD(T)-RKHS PES
(orange), respectively.\\

\noindent
All spectra are normalized such that the maximum peak intensity is set
to unity. As for Ne--H$_2^+$, the QCT simulations accurately reproduce
the experimental peak positions in comparison to the peak
intensities. The positions of the peaks for $j' \in [0,7]$ compare
very favourably between measurements and simulations whereas for $j' =
8$ the energy is somewhat too high. Likewise, the intensity pattern is
correctly captured except for $j' = 4,6$. It is interesting to note
that the CCSD(T)-KerNN PES performs somewhat better for the relative
intensities, in particular for $j' = 3, 4, 8$. It is noted that
quantum full coupled channels calculations the positions and
intensities of {\it all} peaks are realistically
described.\cite{MM.heh2:2023,MM.morph:2024} The feature at ${\rm TKER}
\sim 0$ cm$^{-1}$ arises again from trajectories for which H$_2^+$
($v=v'=1$).  \\

\begin{figure}[h!]
\centering \includegraphics[scale=0.6]{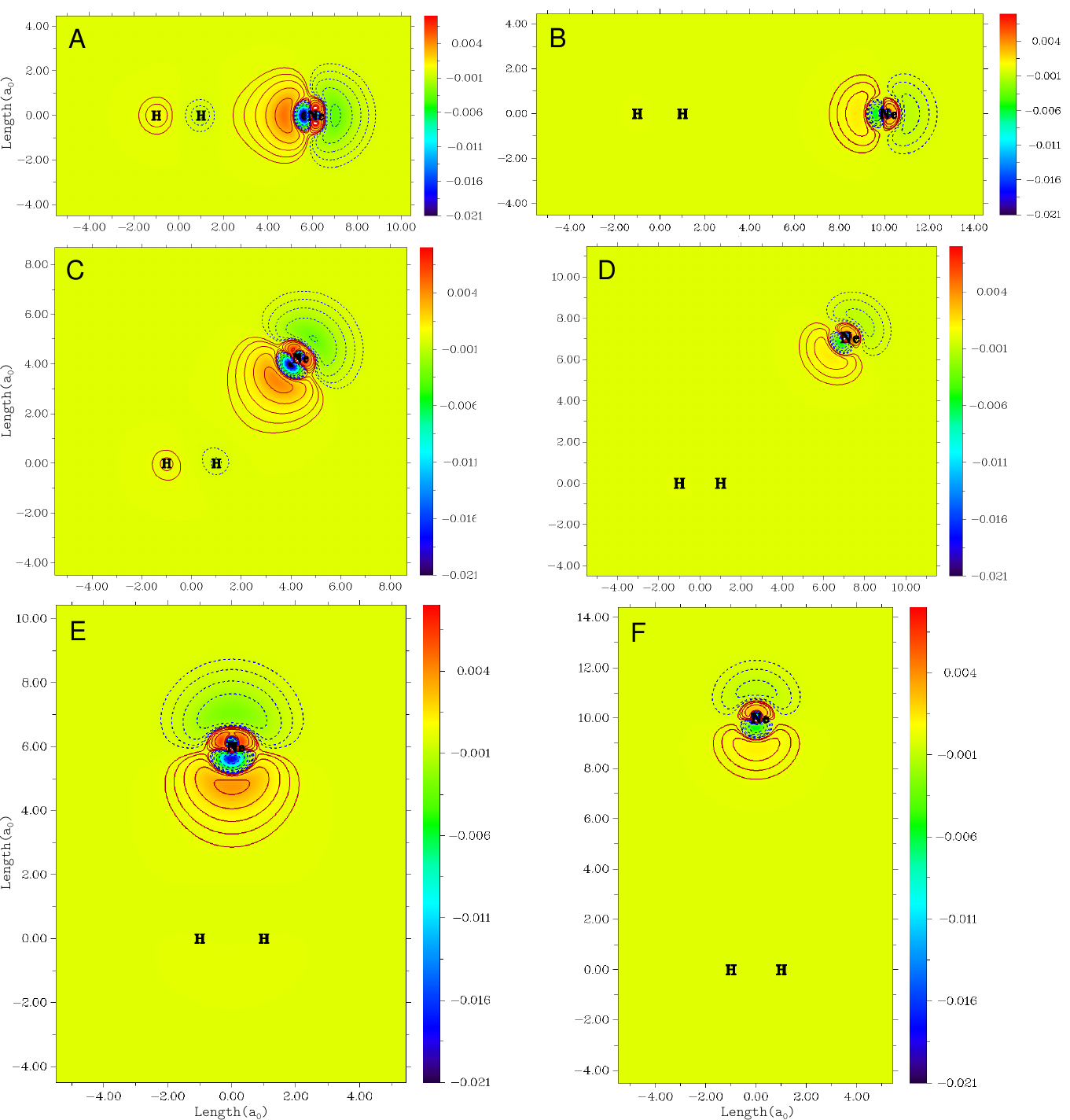}
\caption{Electron density difference maps $\Delta \rho(x,y,z)$ for the
  (Ne--H$_2^+$) complex relative to its isolated components (Ne and
  H$_2^+$), at approach angles of $\theta = 0^\circ$ (A and B),
  $45^\circ$ (C and D), and $90^\circ$ (E and F). Calculations were
  carried out at the CCSD/aug-cc-pVTZ level of theory for
  intermolecular separations of $R = 6\ a_0$ (left panels) and
  $10\ a_0$ (right panels). Solid contours denote regions of increased
  electron density, while dashed contours indicate regions of
  decreased electron density upon complex formation.}
\label{fig:drho-neon}
\end{figure}

\noindent
Since the TKER distributions are particularly sensitive to the
long-range part of the PES, it is also of interest to consider the
intermolecular interactions from a different perspective. For the
Ne--H$_2^+$ complex the electron density difference maps for the two
constituents in the complex compared to the isolated monomers were
determined, see Figure \ref{fig:drho-neon}. This was done for three
different orientations: collinear: $\theta = 0^\circ$ (A/B), $\theta =
45^\circ$ (C/D), and $\theta = 90^\circ$ (E/F). For this analysis the
total electron density $\rho_{\rm tot}(x,y,z)$ from a single point
calculation for fixed geometry of the complex was determined at the
CCSD/aug-cc-pVTZ level of theory. Next, the densities of the isolated
Ne atom $\rho_{\rm Rg}(x,y,z)$ and the H$_2^+$ ion $\rho_{\rm
  ion}(x,y,z)$ in the same geometry as for the calculation of the
complex were determined. Finally, the difference $\Delta \rho(x,y,z) =
\rho_{\rm tot}(x,y,z) - \rho_{\rm Rg}(x,y,z) - \rho_{\rm ion}(x,y,z)$
was determined by superimposing the densities of the isolated rare gas
and ion to the position in the rare gas--ion complex. These difference
densities are shown in Figure \ref{fig:drho-neon} and Figure
\ref{sifig:drho-helium} reports results for He--H$_2^+$\\

\noindent
For Neon, the electron density differences are rather pronounced and
clearly feature an induced dipole originating from the presence of the
H$_2^+$ ion. First, the density difference closer to the ion adopts an
"egg-shape" with a clear elongation towards the ion. Furthermore,
there are more subtle changes in H$_2^+$ electron density closer to
the position of the rare gas atom. As neon moves away from the ion
(compare panels A, C, E with B, D, F), the difference density assumes
more of a "banana shape". Given that the RKHS-based PESs used in the
QCT simulations yield quantitatively accurate TKER spectra when
compared with measurements, indicates that they capture such essential
changes in the electron densities governing the intermolecular
interaction, particularly in the long-range region.\\

\noindent
In summary, the present work demonstrates that QCT simulations provide
a meaningful, informative, and computationally efficient approach to
characterizing the breakup dynamics of Rg--H$_2^+$/HD$^+$. The
calculated TKER distributions quantitatively reproduce measured
positions and relative intensities of most experimentally observed
features. Such speedup and the massively parallel nature of QCT
simulations also provide opportunities for rapid, experiment-guided
improvement of morphed molecular
PESs.\cite{MM.morphing:1999,MM.morph:2024} Furthermore,
trajectory-resolved QCT analysis allows product rotational excitation
to be correlated directly with complex lifetimes. This direct access
to individual dynamical histories offers a practical advantage for
mechanistic interpretation over quantum wavepacket calculations.\\

\section*{Data Availability}
The codes and data for the present study are available from
\url{https://github.com/MMunibas/rgh2} upon publication.

\section*{Acknowledgment}
The authors acknowledge discussions with Prof. C. Koch. Financial
support from the Swiss National Science Foundation through grants
$200020\_219779$ (MM), $200021\_215088$ (MM), and the University of
Basel (MM), the Cluster of Excellence RESOLV (EN) and Alexander von
Humboldt Professorship endowed by the Federal Ministry of Research,
Technology and Space (EN) is gratefully acknowledged. OR acknowledges
funding from Ministerio de Ciencia, Innovaci\'{o}n y Universidades,
MICIU (Spain), under grant No. PID2024-156686NB-I00 \\

\clearpage

\renewcommand{\thepage}{S\arabic{page}}
\renewcommand{\thetable}{S\arabic{table}}
\renewcommand{\thefigure}{S\arabic{figure}}
\renewcommand{\theequation}{S\arabic{equation}}
\renewcommand{\thesection}{S\arabic{section}} 
\setcounter{figure}{0}  
\setcounter{section}{0}  
\setcounter{table}{0}

\section*{Supporting Material}

\section{Experiments}

\begin{center}
\begin{table}[h]
    \centering
\begin{tabular}{||c c c c c c c c c||} 
 \hline
 System & $T_1$  & $v_1$  & $dv_1$  & $T_2$  & $v_2$ & $dv_2$  & $E$  & $dE$ \\ [0.5ex] 
 \hline\hline
 Ne$^*$+HD & 180 & 614 & 10 & 140 & 710 & 16 & 2.19 & 0.33   \\
 \hline
 \end{tabular}
    \caption{Supersonic beam properties: $T$, $v$, $dv$ are the valve
      temperature ([K]), mean velocity ([ms$^{-1}$]) and one standard
      deviation of velocity spread for the Rg (subscript 1) and
      diatomic (subscript 2). $E$ and $dE$ are the corresponding
      collision energy([K]) and its spread. The chemical composition
      of the two beams was pure Ne and a 50/50 Ne/HD mixture,
      respectively.}
    \label{sitab:beam_prop}
\end{table}
\end{center}

\begin{figure}[h]
\centering \includegraphics[width=0.5\textwidth]{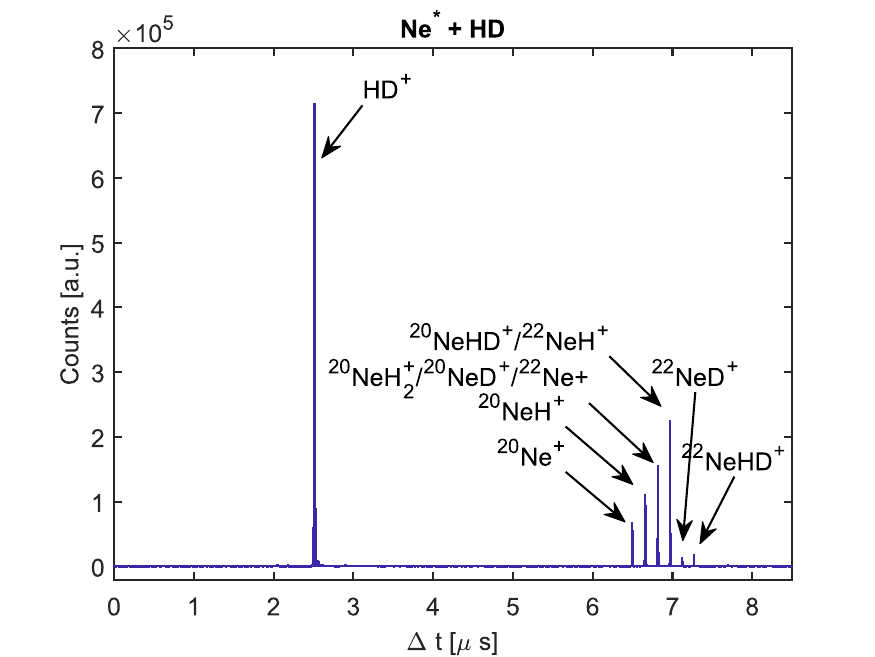}
\caption{Mass spectrum of ionic products after Penning ionization
  collision between Ne$^*$ and HD at 2.19 K. Formation of HD$^+$ and
  other, Ne-containing products is evident.}
\label{sifig:MS_NeHD}
\end{figure}

\begin{figure}[h]
\centering \includegraphics[width=0.5\textwidth]{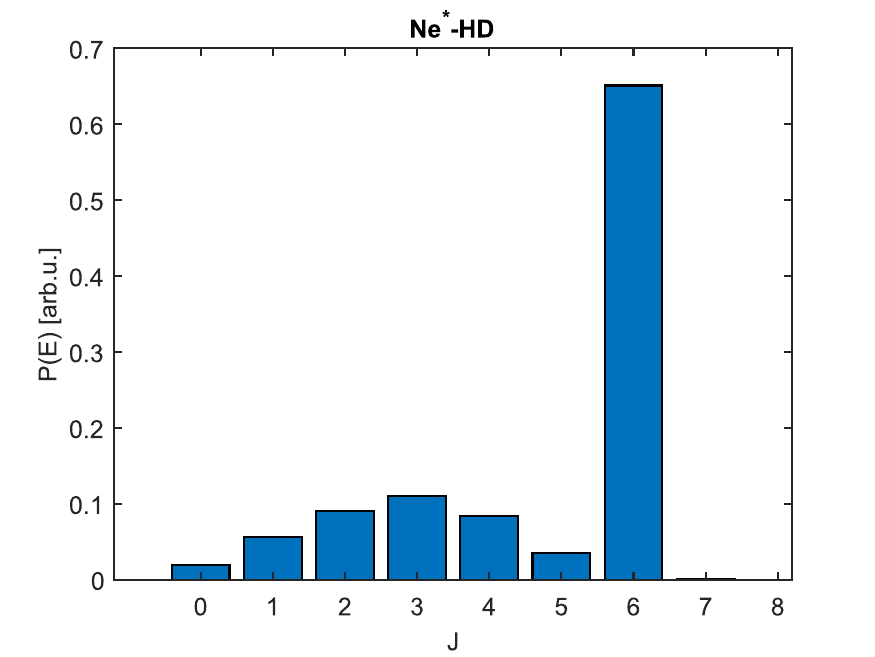}
\caption{Rotational state population with total angular momentum $J =
  l+j$ (partial wave plus molecular rotation) for Ne--HD before
  Penning ionization at 2.19 K. For HD in a supersonic beam, $j=0$ and
  $J=l$.}
\label{sifig:W_NeHD}
\end{figure}

\begin{figure}[h]
\centering \includegraphics[width=0.9\textwidth]{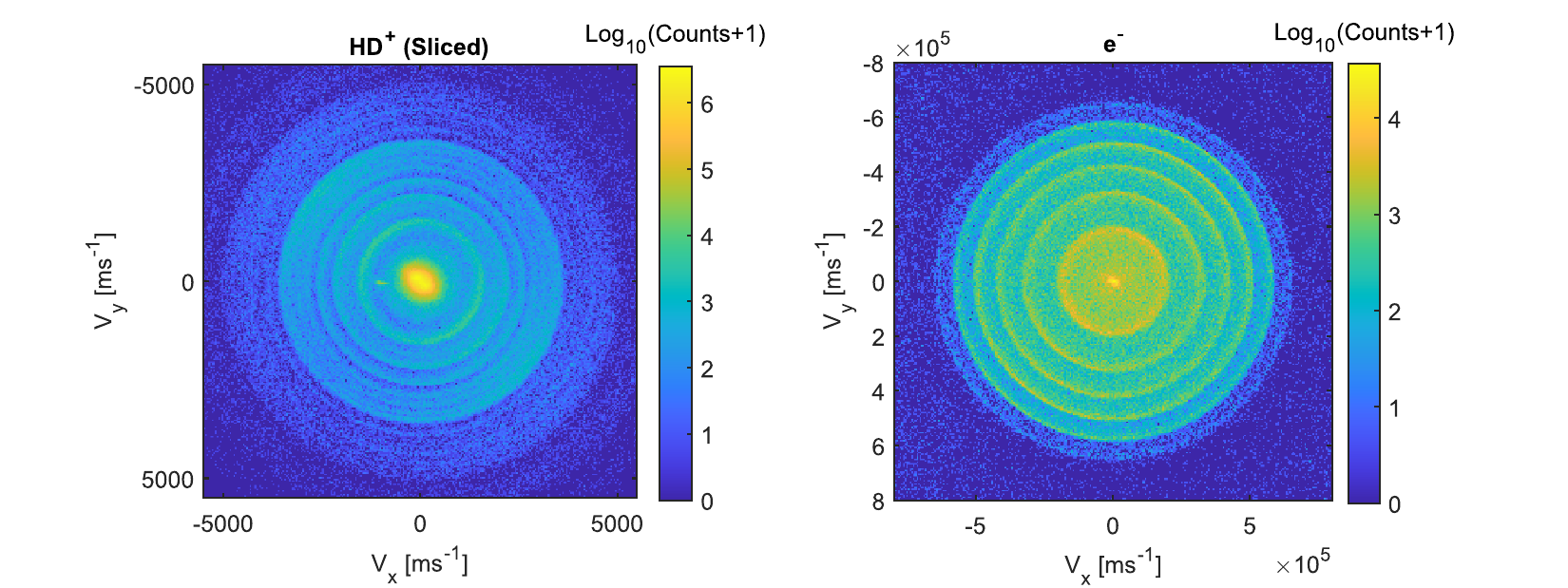}
\caption{Velocity map images of the charged products after Penning
  ionization of of Ne$^*$--HD. Left panel for HD$^+$ and right panel
  for e$^-$ from measurements in coincidence.}
\label{sifig:VMI_NeHD}
\end{figure}

\begin{figure}[h]
\centering \includegraphics[width=0.95\textwidth]{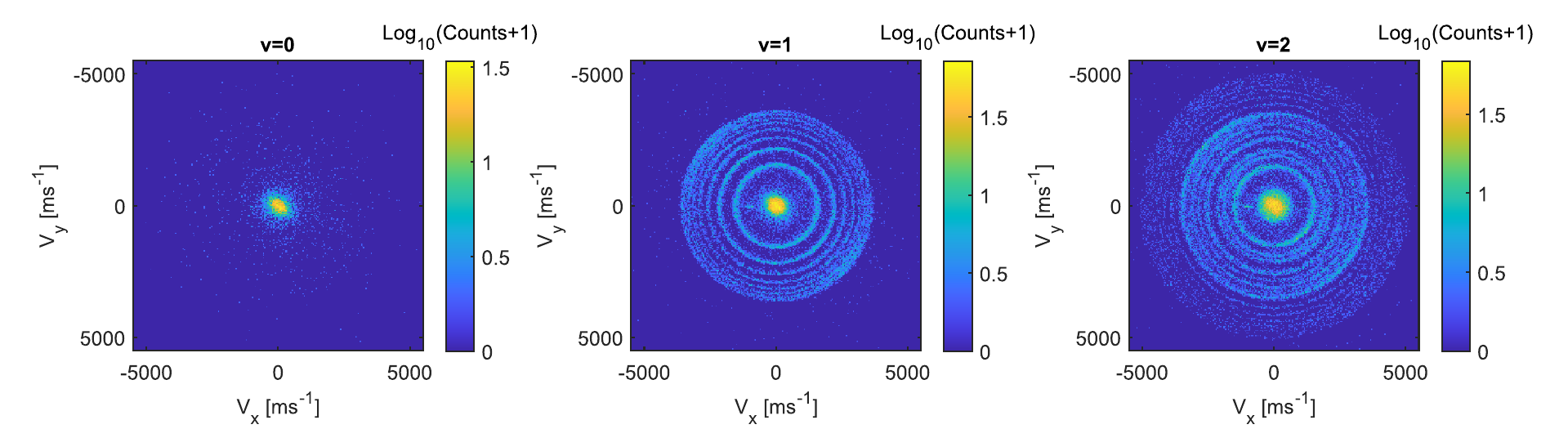}
\caption{Velocity map images for the molecular cations for each
  initial state $v \in [0,2]$ from left to right. Each ring
  corresponds to a final $j'$.}
\label{sifig:VMI_NeHD_per_v}
\end{figure}

\clearpage

\section{Simulations}

\begin{figure}[h!]
    \centering
    \includegraphics[width=0.5\linewidth]{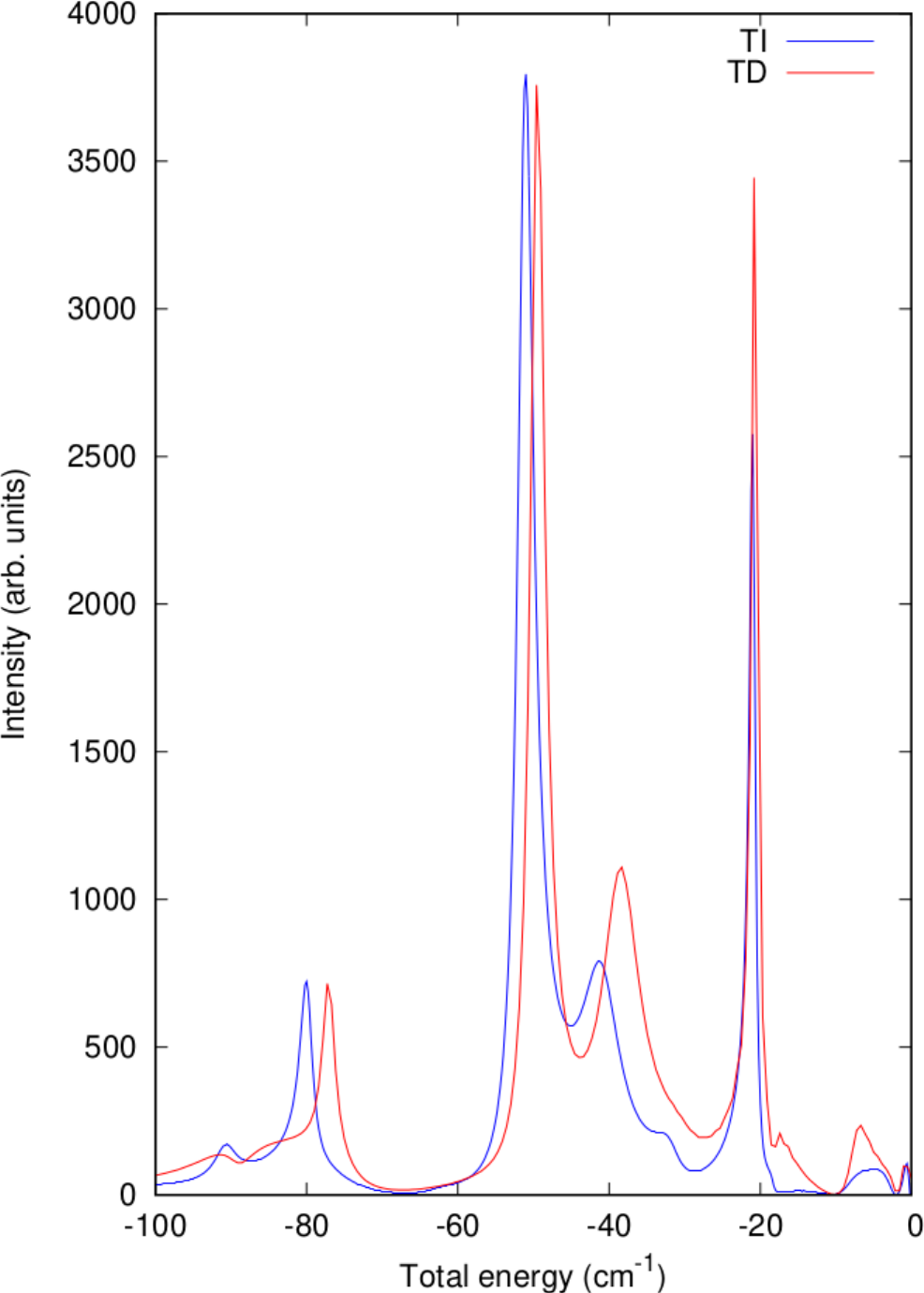}
    \caption{Comparison between time dependent and time independent
      quantum wavepacket simulations (integral cross section as a
      function of total energy) for cold (0.02 K) collisions between
      Ne and HD$^+ (v=1)$ for $J=0$.}
    \label{sifig:compare-wp-j0}
\end{figure}

\begin{figure}[h!]
    \centering
    \includegraphics[width=0.5\linewidth]{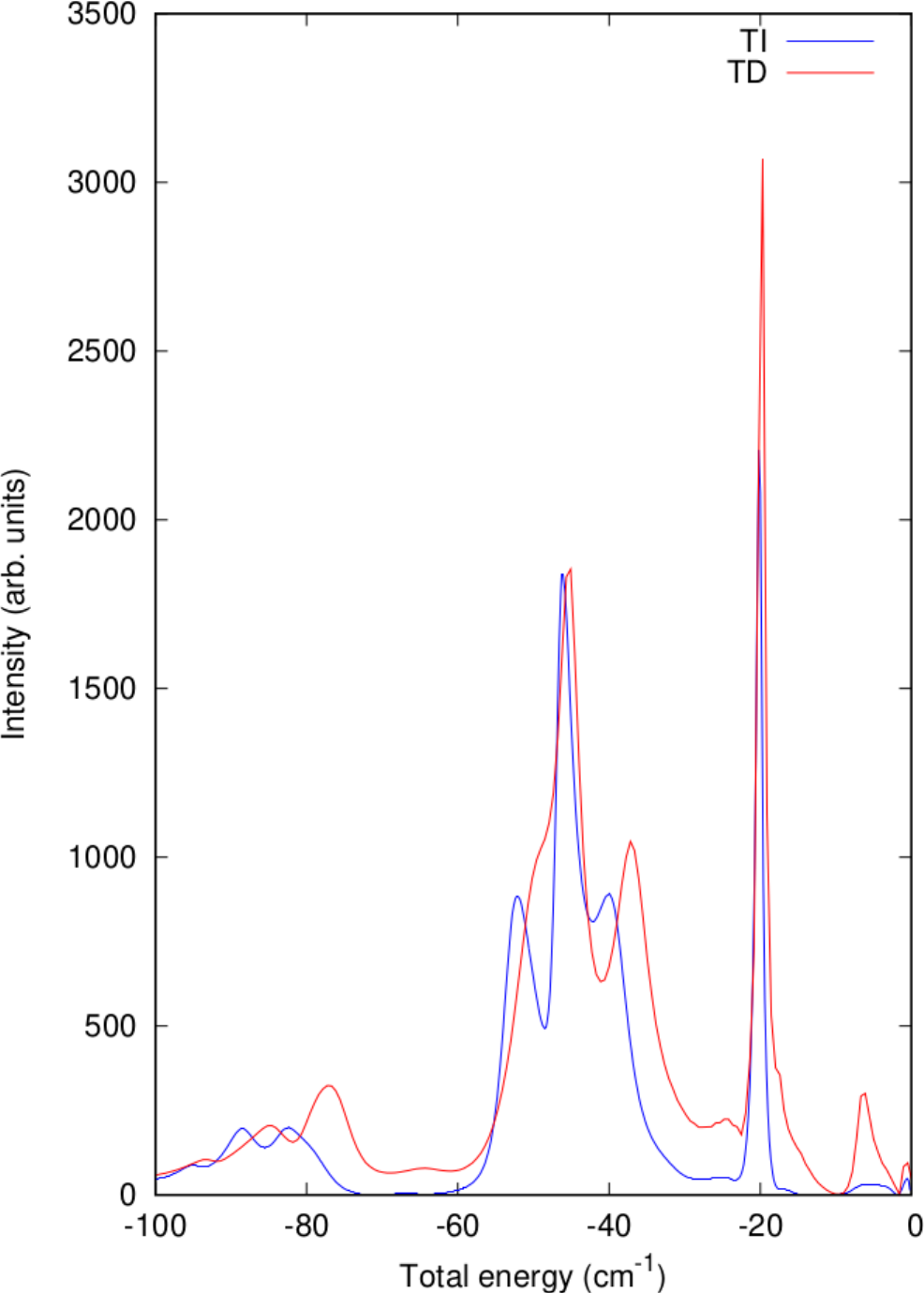}
    \caption{Comparison between time dependent and time independent
      quantum wavepacket simulations (integral cross section as a
      function of total energy) for cold (0.02 K) collisions between
      Ne and HD$^+ (v=1)$ for $J=1$. Due to overlapping resonances
      (highly excited X-HD$^+$ stretching excitations) small
      differences depending on the quantum method used
      arise. Overlapping resonances are very sensitive to details such
      as phases.}
    \label{sifig:compare-wp-j1}
\end{figure}

\begin{figure}[h!]
\centering \includegraphics[scale=0.6]{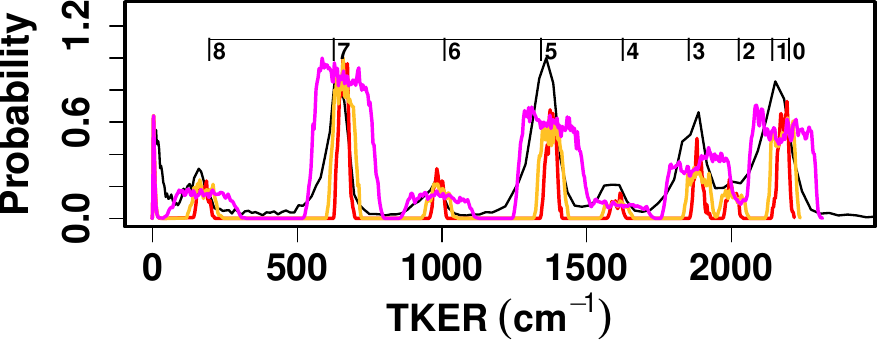}
\caption{Comparison of the TKER spectra calculated using binning
  tolerances of 0.01 (red), 0.02 (golden) and 0.05 (fuchsia) for Ne +
  H$_2^+ (v=1, j=0/1; J\in[4,6])$ $\longrightarrow$ Ne + H$_2^+ (v'=0,
  j')$. The experimental spectrum is shown in black.}
\label{sifig:neh2p-difftol}
\end{figure}

\begin{figure}
    \centering \includegraphics[scale=1.7]{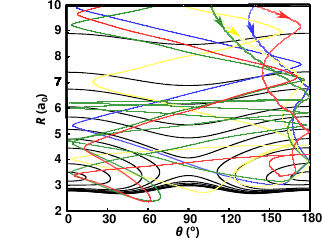}
    \caption{Mapping of 4 vibrationally relaxing (in blue, red, green,
      and brown color solid lines) trajectories $(v=1, j=0)
      \longrightarrow (v'=0, j'=9)$ for Ne + HD$^+$ collisions with
      initial conditions $J=6, v=1, j=0$, $R_{\rm 0}= 10$ a$_0$.}
       \label{sifig:nehdp-mapping}
\end{figure}

\clearpage

\begin{figure}[H]
\centering \includegraphics[scale=0.6]{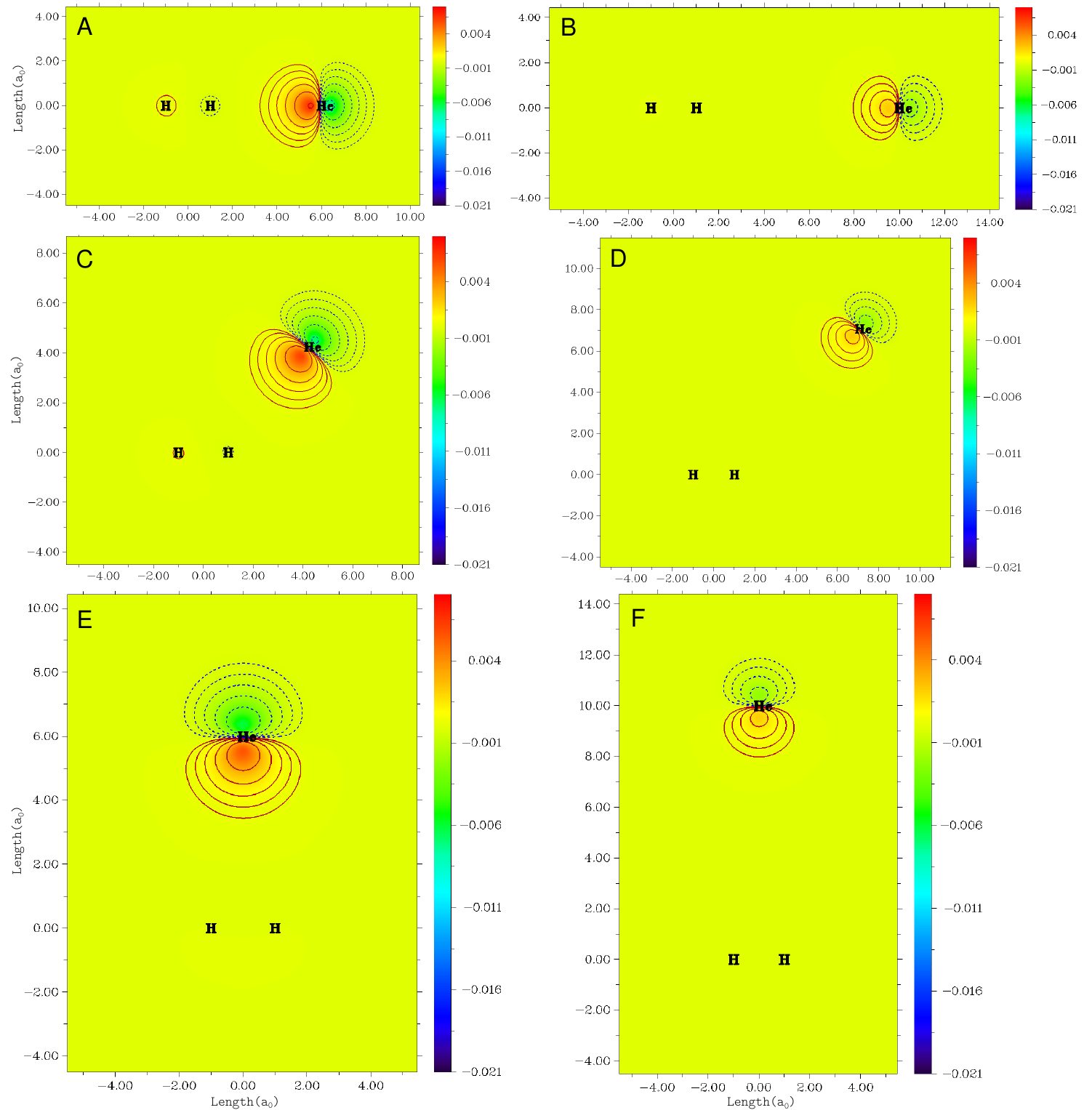}
\caption{Electron density difference maps for the (He---H$_2^+$)
  complex relative to its isolated components (He and H$_2^+$), at
  approach angles of $\theta = 0^\circ$ (A and B), $45^\circ$ (C and
  D), and $90^\circ$ (E and F). Calculations were carried out at the
  CCSD/aug-cc-pVTZ level of theory for intermolecular separations of
  $R = 6\ a_0$ (left panels) and $10\ a_0$ (right panels). Solid
  contours denote regions of increased electron density, while dashed
  contours indicate regions of decreased electron density upon complex
  formation.}
\label{sifig:drho-helium}
\end{figure}

\clearpage

\bibliography{refs}

\end{document}